# Defects-Assisted Piezoelectric Response in Liquid Exfoliated MoS$_2$ Nanosheets

Jyoti Shakya[1], Gayathri H N[1], and Arindam Ghosh[1,2]

[1]Department of Physics, Indian Institute of Science, Bangalore 560012, India

[2]Centre for Nano Science and Engineering, Indian Institute of Science, Bangalore 560012, India

We report piezoelectric response in liquid phase exfoliated MoS$_2$ nanosheets with desired structure and morphology. The piezoelectric effect in liquid phase exfoliated few layers of MoS$_2$ flakes is interesting as it may allow the scalable fabrication of electronic devices such as self-powered electronics, piezoelectric transformers, antennas and more. The piezo force microscopy (PFM) measurements were used to quantify the amplitude and phase loop, which shows strong piezoelectric coefficient. Herein, the piezoelectric response in few layers of MoS$_2$ is attributed to the defects formed in it during the synthesis procedure. The presence of defects is confirmed by XPS analysis.

Piezoelectricity is the charge created when certain materials are placed under stress. The compressing or stretching the substance generates electricity and allow facile conversion of mechanical energy into electrical and vice versa which offers immense application for sensors, energy harvesting, flexible electronics, energy storage and actuators. Great deal of research is being carried out in generating clean energy due to obvious reasons.[1-2] As it is essential to find non-conventional forms of energy which are sustainable and environment friendly, harvesting electricity from human surrounding is being considered as one way to substitute conventional fuels. Mechanical energy can exist in different forms such as vibration, wind, sound wave, flow of water and energy from human body movement and so on, which are thought to be useful sources to harness clean energy. The piezo electric systems could prove extremely useful in such scenario. Piezoelectricity at micro and nanoscale can be effectively engineered by necessary surface modifications including the introduction of adatoms and defects in such materials leading to surface piezoelectricity. Vast amount of research has been reported on piezoelectricity in different inorganic and organic based materials including composites.[3-5] However, some of the above-mentioned piezoelectric materials have some drawbacks because of their toxic nature, non-biodegradability, rigidity, non-ecofriendliness, non-biocompatibility, and lower piezoelectric properties. In search of efficient piezoelectric material, 2D materials are emerging as a potential alternative and they have been theoretically predicted to be piezoelectric and some of them have been confirmed experimentally. The transition metal dichalcogenides (TMDCs) in particular have been explored increasingly due to their non-centrosymmetric crystal structure and large surface area to thickness ratio.[6-8] As a typical TMDC material, $MoS_2$ is of great interest due to its novel properties such as tunable band gap, large in-plane mobility and mechanical stability.[9-12] Each layer of $MoS_2$ consists of molybdenum atoms sandwiched between two layers of hexagonally close-packed sulfur atoms, and the weak van der Waals forces tie the sandwiched adjacent layers.[13] Ultrathin $MoS_2$ layers can be prepared through various techniques such as hydrothermal method, liquid exfoliation, chemical vapor deposition, mechanical exfoliation and so on.[14-17] The technique of liquid phase exfoliation looks promising for scalable production of $MoS_2$ because of its low cost, relatively high output and simplicity. But controlled synthesis of crystalline $MoS_2$ with desired structure and morphology is still a challenge.

In case of $MoS_2$ it has been observed that the odd layered $MoS_2$ is highly piezoelectric due to non-centrosymmetric nature, with monolayer showing the strong piezoelectricity.[6, 20] The piezoelectric effect disappears in bulk $MoS_2$ due to the opposite orientations of the adjacent layers. Most of these demonstrations of piezoelectric effect have been made on mechanically exfoliated odd number of nanosheets which pose limitation for large scale applications. As an alternative, the technique of liquid phase exfoliation (LPE) along with fabrication techniques like solution casting, spray coating, or inkjet printing can be thought of for scalable fabrication of electronic devices. In this direction, some theoretical research on piezoresponse in multilayer 2D materials have been reported but intensive experimental studies are still lacking.[6, 8, 21] To the best of our knowledge, no experimental report is available on the piezoelectric response from liquid phase exfoliated few layers of $MoS_2$. In this work, we propose that 2D $MoS_2$ layers synthesized via modified LPE method contain defects which can act as prime spots for trapping charges and display piezo signal.

$MoS_2$ nanosheets have been synthesized using modified liquid phase exfoliation method. In detail, 2g of $MoS_2$ (powder, < 6µm, 98%, Sigma) was dispersed in 60 ml water and probe sonicated for 1h to get a homogeneous mixture. The obtained mixture was subjected to hydrothermal reaction with continuous stirring for 3h at 150 ºC. The pre-treated $MoS_2$ was further exfoliated optomechanically for about 5 hours and the mixture was centrifuged at 5000 rpm for 30 minutes. The supernatant with concentration of 20 µg/L was deposited on ITO substrate by spin coating, followed by thermal annealing at 100 ºC for about 60 minutes.

Different techniques have been used for the characterization of synthesized few layers of $MoS_2$. The surface morphology of $MoS_2$ nanosheets was observed and analysed by scanning electron microscope coupled with energy dispersive X-ray (SEM-EDX) spectroscopy (Ultra55 FE-SEM Karl Zeiss EDS by Karl Zeiss). All the SEM images were taken under the operating voltage of 5kV in InLens mode. The qualitative analysis was carried out using energy dispersing X-ray (EDX) analysis at electron high tension (EHT) voltage of 15 kV. The Raman spectra were taken at room temperature using Horiba LabRAM HR Raman spectrometer. Instrument resolution was ~ 0.5 $cm^{-1}$ by using 1800 groove/mm grating. A linearly polarized laser light (532 nm) was used for the measurements keeping the laser power below 1 mW. It was focused to a spot of 1 µm diameter by a 50x long working distance objective lens. Structural analysis and mapping were carried out using a transmission electron microscope (TEM TITEN Themis) conducted at accelerating voltage of 200 kV. TEM analysis was made by drop-casting the diluted $MoS_2$ dispersion over the carbon-coated copper grid, followed by drying under IR bulb. The PFM measurements were carried out using atomic force microscope, Park system (Park AFM NX 20). Conducting cantilever (Multi75E-G) having force constant 3 N/m and resonant frequency 75 kHz was employed in all our PFM measurements.

The crystallinity of $MoS_2$ layers was analysed by TEM. Sheets like structure of $MoS_2$ and its HRTEM image are presented in Fig. 1(a) and Fig. 1(b) respectively. A clear view of lattice fringes is shown in Fig. 1(b) and the lattice spacing of the order of 0.27 nm which corresponds to (100) plane of $MoS_2$ nanosheets, can be observed.[10] The inset of figure 1(b) shows the SEAD pattern of $MoS_2$ confirming the crystallinity of the exfoliated nanosheets. The $MoS_2$ sheets in figure 1(a) suggest the presence of very thin sheets. The crystal structure was also confirmed by Raman spectroscopy (fig 1c). The peak difference of ~ 25 $cm^{-1}$ between the characteristic peaks of MoS$_2$ confirms the presence of few layers.[22] The photoluminescence spectra of MoS$_2$ layers is shown in fig 1 (d). The Raman and PL spectra also shows the characteristic peaks corresponding to MoS$_2$ few layers.

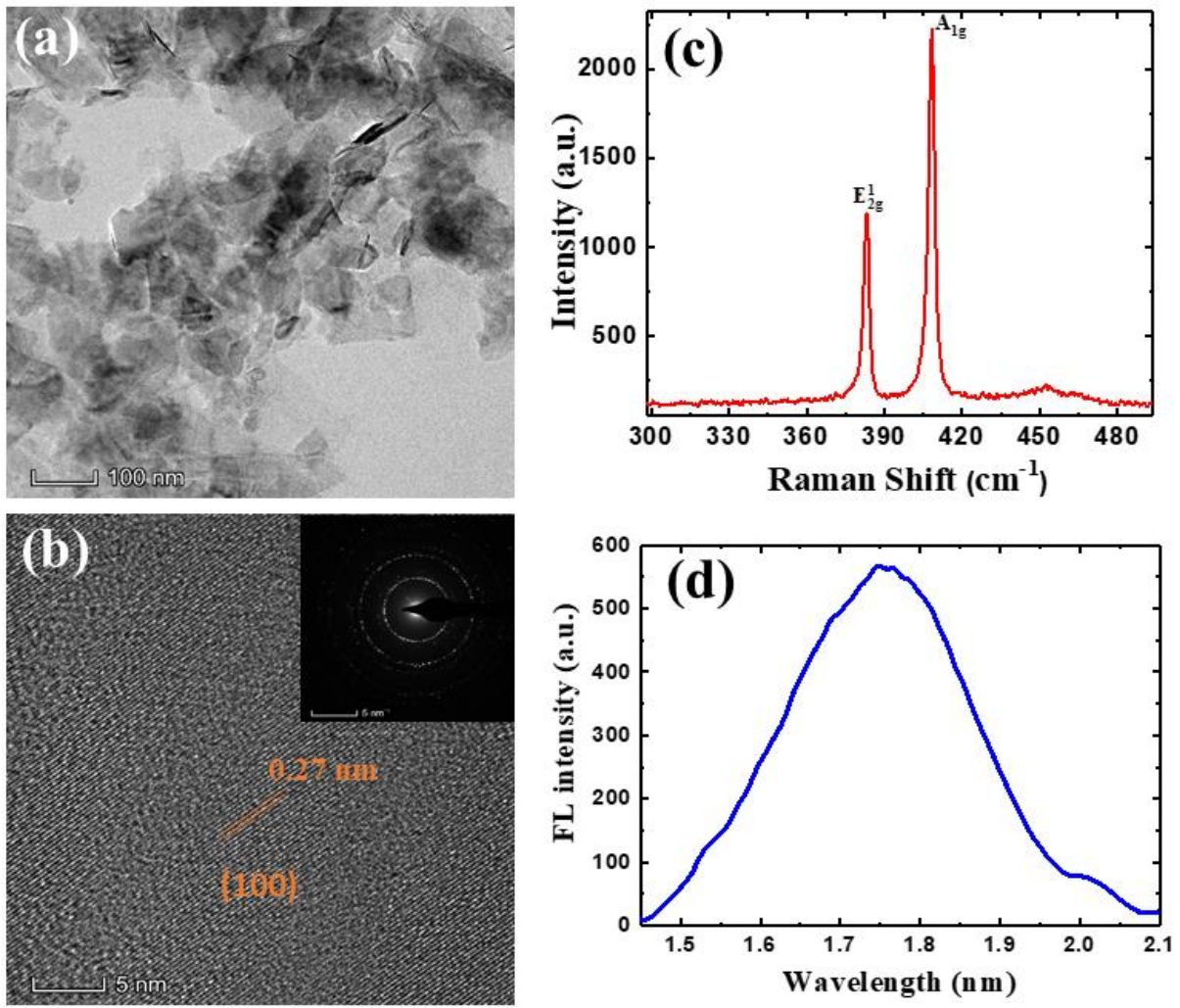

FIG. 1. (a) TEM and (b) HRTEM images with inset 1(b) showing the SAED pattern of MoS$_2$ nanosheets. (c) Raman spectra and (d) Photoluminiscence spectra of few layers MoS$_2$.

The SEM image in figure 2a shows exfoliated nanosheets and the corresponding lateral area distribution of the flakes are shown in figure 2b. The average flake area is found to be

0.4663µm². The figure 2c shows the EDAX spectra of MoS$_2$ nanosheets and the spectra confirms that the as-synthesized MoS$_2$ nanostructures contain Mo and S elements along with oxygen. The thickness distribution of exfoliated nanosheets, analysed using AFM had average flakes thickness around 2.96 nm (figure 2d and 2e). This suggested that bulk MoS$_2$ particles are successfully exfoliated and significantly fragmented. These are not absolute limits simply because such a small sample is extremely unlikely to contain the largest and smallest flakes in the distribution. Therefore, a small population of flakes outside these limits probably exist although we did not observe them during our statistical analysis.

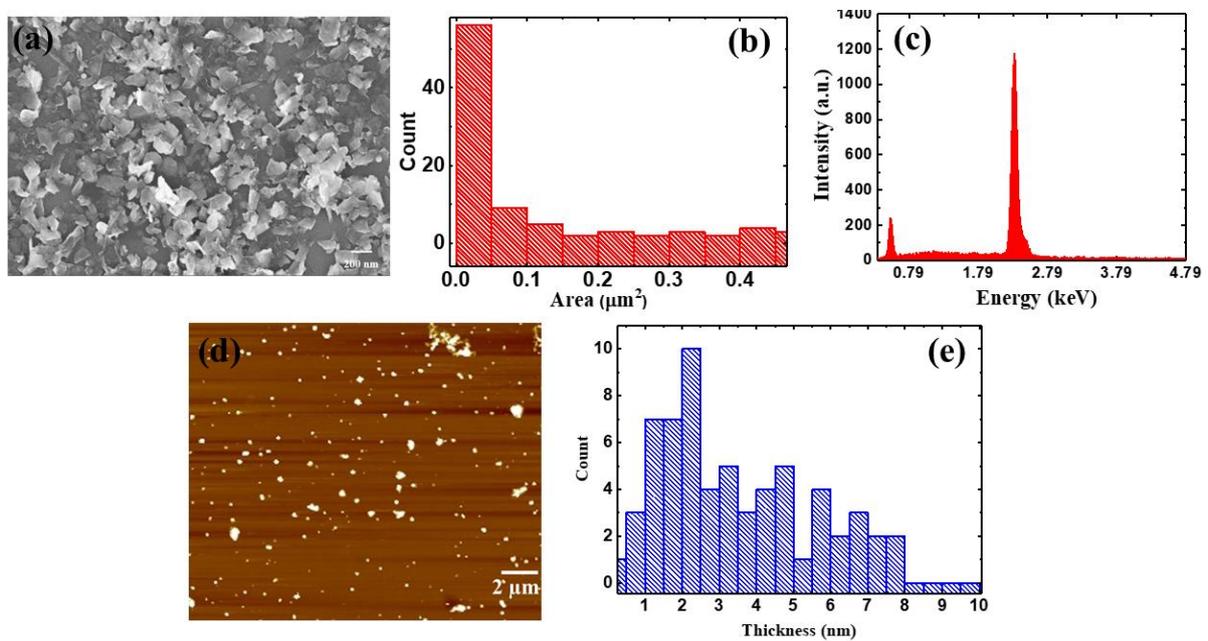

FIG. 2. Nanosheet distribution shown by SEM and AFM images. (a) SEM image (b) nanosheets lateral area distribution, (c) EDS spectra (d) AFM image and (e) nanosheet thickness distribution of MoS$_2$ nanosheets.

To quantify the piezoelectric response at nanoscale, Piezo force microscopy (PFM) can be used. The PFM based investigations were carried out using AFM in contact mode. The PFM uses the basic principle of inverse piezoelectric effect: AC field is applied onto the sample surface through AFM probe keeping the bottom electrode grounded, which results in the surface deformation of the sample and the amplitude/phase of this precise deformation could be measured to characterize the strength/direction of the dipole. The tip-sample voltage is modulated with a periodic tip bias $V_{tip} = V_{DC}+V_{AC}\cos(\omega t)$, here the drive frequency $\omega$ is chosen to be well above the feedback bandwidth. This drive is responsible for generating an oscillating electric field below the tip that causes localized deformation in the sample surface. During the measurement, the cantilever oscillation frequency was directed close to its resonant frequency with $V_{bias}$ = 6 V to acquire the piezoelectric coefficient ($d_{33}$) values at 50 kHz, which is well below both the free space resonance frequency and the resonance of cantilever with sample in contact.

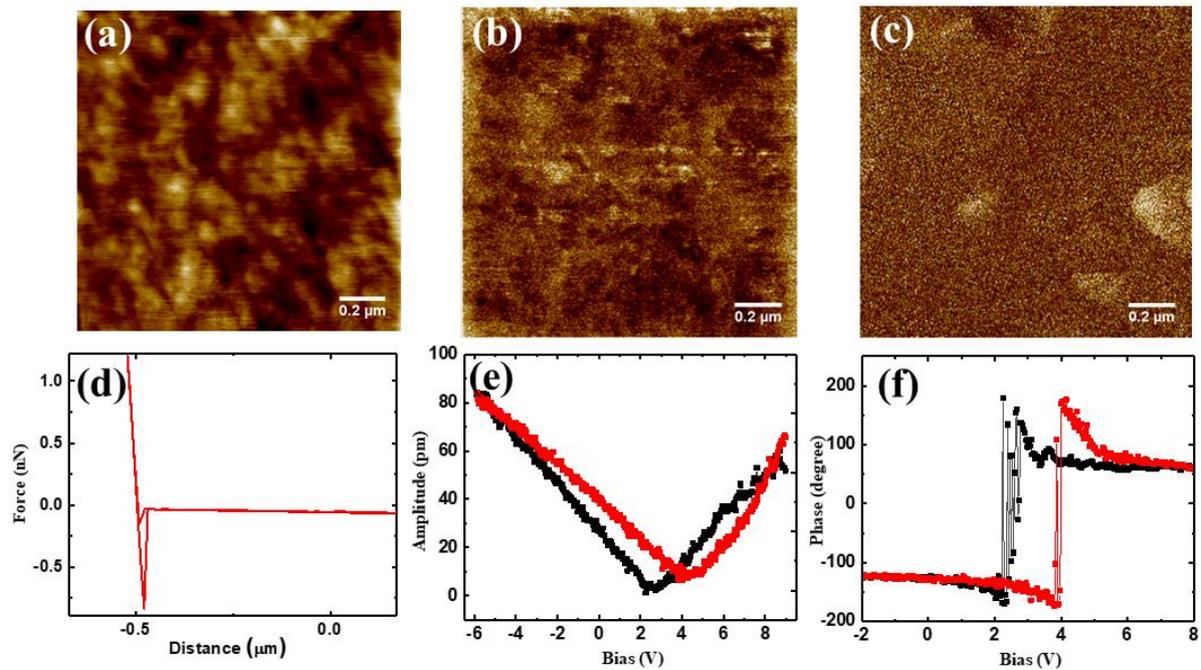

FIG 3. (a) Topography (b) Amplitude (c) PFM phase image of few layer $MoS_2$ nanosheets (d) Force distance curve. (e) PFM amplitude and (f) PFM phase hysteresis loop on few layers of $MoS_2$.

It is observed that there is less than 20% variation in the PFM signal amplitude in the frequency range of 20-60 kHz, and we worked with 50 kHz to optimize the signal stability and signal-to-noise ratio.

To begin with, the PFM measurements were carried out on few layers of $MoS_2$. The figures 3a, 3b and 3c show the topography, amplitude and phase image of $MoS_2$ layers. The tip (Bruker SCM-PIT) of AFM acts as top electrode during PFM measurements. The deflection sensitivity of the system (48 V/µm) was determined using force-distance (F-D) curve (figure 3d) by indenting the cantilever tip on a hard surface of sapphire. To get electrostatic free piezoresponse, the voltage was applied during PFM measurements and a stiffer (2.945 N/m) cantilever was used throughout the measurement. The PFM response is quantified in terms of its amplitude (*A*) and phase (φ) signals, by comparing the input voltage with periodic surface vibrations using a lock-in amplifier. The characteristic butterfly loop in the PFM amplitude and the hysteresis loop in the phase signal as a function of applied voltage for few layers of $MoS_2$ are also shown in figure 3 (e) and 3 (f) respectively. In order to determine the value of piezoelectric constant, the piezoresponse was measured by different driving voltages repeatedly. The change in PFM amplitude indicates mechanical responses with respect to the electrical signal of $V_{tip}$ confirming the piezoelectric effect from $MoS_2$ nanosheets. To quantify the effect, the linear piezoelectricity can be determined by the following equation,[23]

$$A = d_{33} V_{ac} \quad\quad\quad (1)$$

where, *A* is the PFM amplitude, $d_{33}$ is the effective piezoelectric coefficient and $V_{ac}$ is the drive ac bias voltage. The measurements were conducted in the strong-indentation regime, where the piezo-response is dominated by the $d_{33}$ rather than $d_{31}$ of the material. Here $d_{33}$ is the induced strain in direction 3 per unit electric field applied in direction 3 whereas $d_{31}$ is the induced strain in direction 1 per unit electric field applied in direction 3. From the plot of piezo-response amplitude vs. $V_{bias}$, taken on the MoS$_2$ layers (Figure 3e), we extracted the effective piezoelectric coefficient value (d$_{33}$) of 9.6 pm/V.

To confirm that this response was not coming only because of the classical piezoelectricity of MoS$_2$ in odd number of layers, the piezo response of MoS$_2$ having different thicknesses (monolayer, bilayer and tri-layer) was investigated by PFM under similar imaging parameters. All the PFM measurements were carried out in the ramping voltage range of -10V to +10 V for different layers of MoS$_2$ and are shown in figure 4. The AFM topography of MoS$_2$ sample, consisting of mono, bi and tri layer of MoS$_2$ are marked as 1, 2 and 3 respectively in figure 4a. The PFM amplitude signal as a function of ramped $V_{bias}$ at frequency 50 kHz for respective mono-, bi- and tri- layer of MoS$_2$ are shown in figures 4b, 4c and 4d. To extract the piezoelectric coefficient, different ac voltages were applied from the AFM tip, which shows a linear relationship between mechanical deformation and the applied electric field. As expected, a strong piezoelectric response was observed from monolayer MoS$_2$. From the piezo-response amplitude vs. $V_{bias}$ curves taken on the spots marked by the plus signs (Figure 4a), we extracted the effective piezoelectric coefficient ($d_{33}$) value of 10.91 pm/V, 15.75 pm/V and 25.14 pm/V for monolayer, bilayer and tri-layer respectively. The material with a higher piezoelectric coefficient produces more electricity for a given strain thus improving the efficiency of the piezoelectric device.

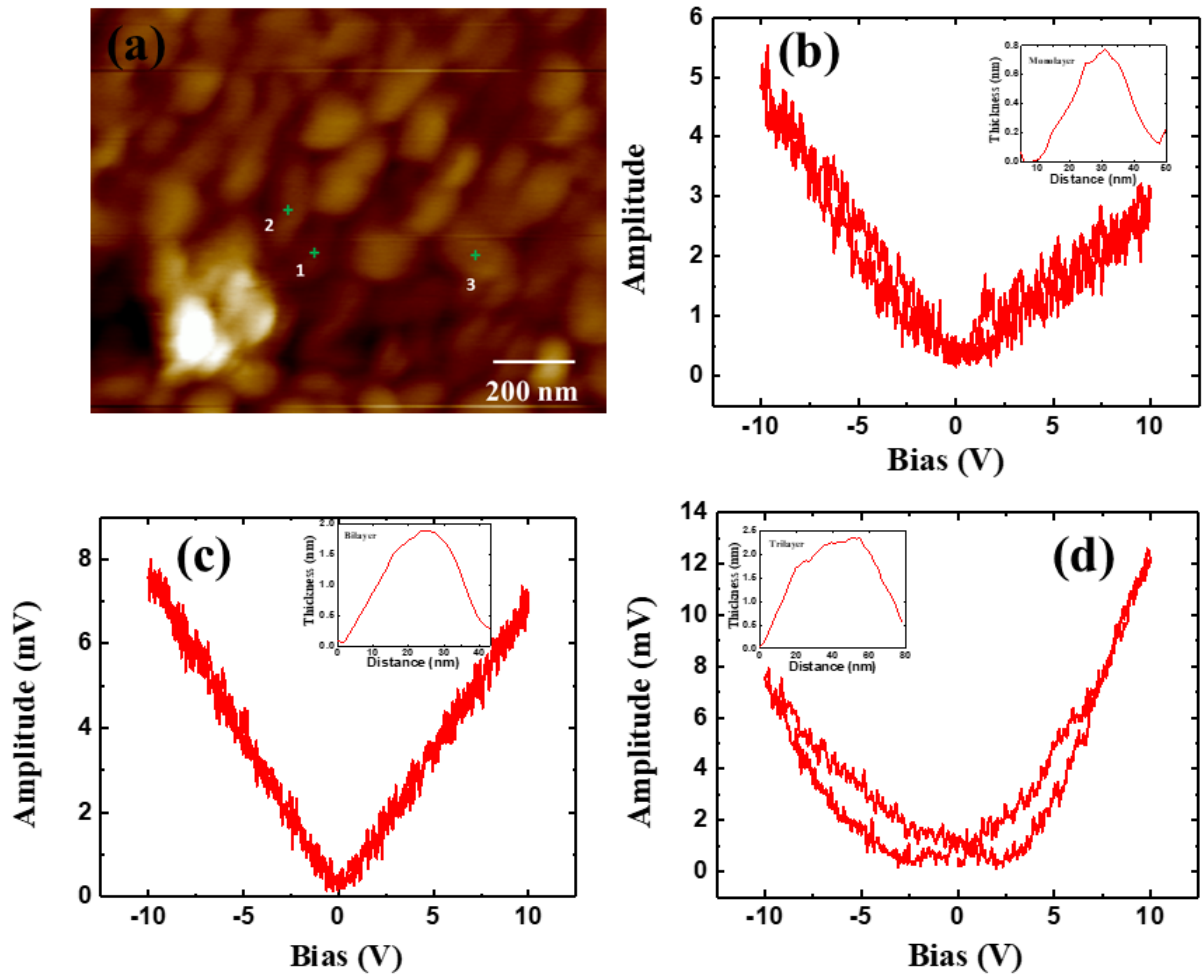

FIG. 4. (a) Topographic AFM image showing mono, bi and tri layer MoS$_2$. PFM amplitude of (b) monolayer (c) bilayer and (d) trilayer MoS$_2$. The inset shows the thickness of the MoS$_2$ nanosheets from topographic image shown in (a) on which the PFM measurements were made.

In general, the piezoelectric effect appears when the atomic symmetry of a material is broken due to an external mechanical stress, which causes accumulation of electric charges. Therefore, the piezo response has been observed in odd number of layers of MoS$_2$.[6] The piezoelectric response from MoS$_2$ in this work is probably coming because of defects as observed through EDX analysis. There is a possibility of trapping and stabilization of charges around the defects in the material as reported by A. Apte et[24] al that can induce piezoelectric response. The synthesis procedure in this work gives rise to oxidation which suggest a route to artificially engineer piezoelectricity in 2D crystals. To further verify the presence of defects, XPS measurements have been performed. In the full survey spectra shown in fig 5a, the characteristic peaks of Mo, S, O and Si (due to substrate) atoms were observed. Except C1s peak, no other impurity peak can be found. The corresponding binding energy plots for Mo 3d and S 2p are presented in figure 5a and 5b. All spectra were referenced by adventitious carbon 1S peak at 284.8 eV (not shown).

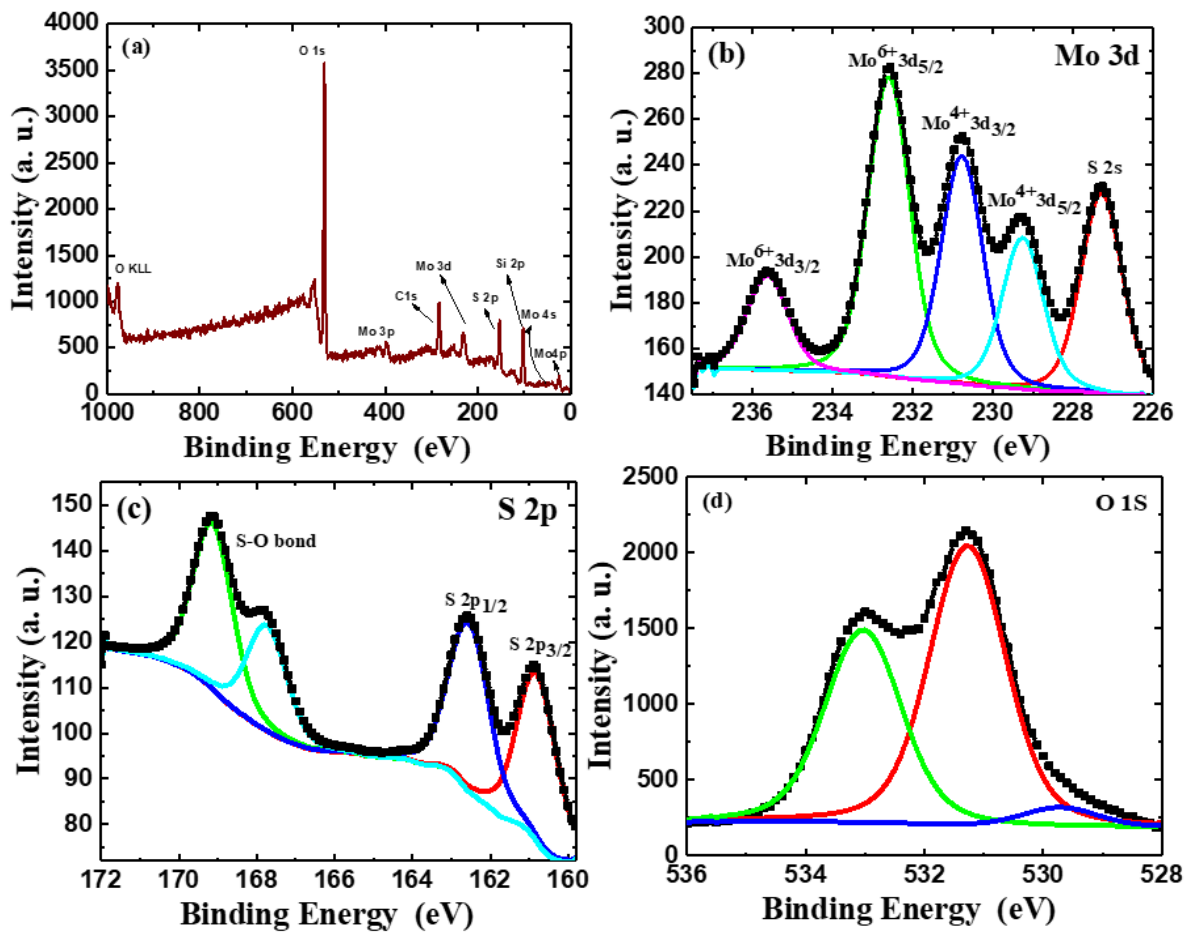

FIG. 5. Photoelectron spectroscopy (XPS) spectra of $MoS_2$ surface (a) full survey spectrum. The high-resolution scanning of the (b) Mo 3d (c) S 2p and O 1s peaks.

In the Mo 3d spectra of as prepared $MoS_2$ layers, the characteristic peaks at 229.25 eV and 231.41 eV belongs to the $Mo^{4+}$ $3d_{5/2}$ and $Mo^{4+}$ $3d_{3/2}$ components of $MoS_2$, respectively.[25, 26] The peak at 227.27 eV arises due to S 2s of $MoS_2$.[25] The binding energies of Mo 3d and S 2p regions reveal that the few layer MoS$_2$ has a strong trigonal prismatic structure. The peak at binding energies 232.59 eV and 235.6 eV are ascribed to Mo$^{6+}$ revealing that the Mo edges in MoS$_2$ layers are slightly oxidized during the transition from the Mo$^{4+}$ state to the Mo$^{6+}$ state which corresponds to the Mo(VI) species (Mo$^{6+}$ 3d$_{5/2}$ and Mo$^{6+}$ 3d$_{3/2}$) in MoO$_x$.[27, 28] Therefore peaks attributes to the formation of Mo-O suggesting some degree of MoO$_x$-like structures. Oxygen can be attached as substitutional atoms at sulfur sites or as atoms bound to Mo atoms at plane edges. Additionally, there are two strong peaks at 162.07 eV and 163.3 eV which are assigned to S 2p$_{3/2}$ and S 2p$_{3/2}$ binding energies for S$^{2-}$ and the peak at 169.2 eV could be ascribed to existence of S-O bond[29] which indicates partial oxidation of the S edges in MoS$_2$ layers. The oxidation percentage of Mo and S edges were 12.04% and 27.13% respectively. The trapping and stabilization of charges around the slight oxidation in MoS$_2$ induced during synthesis process may result into the piezoelectric response even in the few layers of MoS$_2$.[24]

In summary, a strong piezoelectric coefficient in few layers of MoS$_2$ has been observed. The value of piezoelectric constant in MoS$_2$ increases with the increase in its thickness. However, its value shows slight decrease in few layered MoS$_2$. The piezoelectric response in liquid exfoliated layers of MoS$_2$ is attributed to the formation of defects during its synthesis. Given the materials properties and piezoelectric response, liquid exfoliated MoS$_2$ could be considered promising for piezotronics devices and energy harvesting technology.